\documentclass[prd, twocolumn, nofootinbib, floatfix]{revtex4-1}

\usepackage{amsmath}
\usepackage{graphicx}
\usepackage{dcolumn}
\usepackage{bm}
\usepackage{enumerate} 
\usepackage{epsfig}
\usepackage{amssymb,latexsym,mathrsfs}
\usepackage{graphicx}
\usepackage{color}
\usepackage{hyperref}

\hypersetup{
    colorlinks=true,
    linkcolor=red,
    citecolor=blue,
} 

\newcommand{\be}{\begin{equation}}
\newcommand{\ee}{\end{equation}}
\newcommand{\bes}{\begin{equation*}}
\newcommand{\ees}{\end{equation*}}
\newcommand{\beq}{\begin{equation}}
\newcommand{\eeq}{\end{equation}}
\newcommand{\bs}{\begin{split}} 
\newcommand{\bea}{\begin{eqnarray}}
\newcommand{\eea}{\end{eqnarray}}
\newcommand{\beqa}{\begin{eqnarray}}
\newcommand{\eeqa}{\end{eqnarray}}

\newcommand{\al}{\alpha}

\begin{document}

\title{Dark Energy from $\alpha$-Attractors} 
\author{Eric V.\ Linder} 
\affiliation{Berkeley Center for Cosmological Physics \& Berkeley Lab, 
University of California, Berkeley, CA 94720, USA} 

\begin{abstract}
A class of inflation theories called $\al$-attractors has been 
investigated recently with interesting properties interpolating between 
quadratic potentials, the Starobinsky model, and an attractor limit. 
Here we examine their 
use for late time cosmic acceleration. We generalize the class and 
demonstrate how it can interpolate between thawing and freezing dark 
energy, and reduce the fine tuning of initial conditions, allowing 
$w\approx-1$ for a prolonged period or as a de Sitter attractor. 
\end{abstract}

\date{\today} 

\maketitle

\section{Introduction} 

Scalar fields play a fundamental role in cosmology, with the Higgs field 
in the standard model of particle physics, the inflaton seeding early 
universe quantum perturbations that develop into cosmic structure and a 
primordial gravitational wave background, and possibly a dark energy field 
that is responsible for late time cosmic acceleration. One of the exciting 
recent developments in inflation is the discovery of classes of theories 
that connect different scalar field theories. Here we examine aspects 
of this in the context of late time dark energy. 

The inflationary developments involve conformal $\xi$-attractors and 
K{\"a}hler curvature $\alpha$-attractors. These have the property of 
interpolating between power law scalar field potentials or, e.g., Higgs 
inflation, Starobinsky inflation, and a fixed point 
\cite{13065220,13110472,14123797}. Thus such theories make definite 
predictions for the primordial scalar perturbation tilt $n_s$ and the tensor 
(gravitational wave) to scalar power ratio $r$, both measurable by cosmic 
microwave background experiments, with all models approaching 
a particular limit. 

We focus here on the minimally coupled $\al$-attractors and investigate 
the dark energy properties of such models. Several of the forms within 
this class have high energy physics motivations, from supergravity and 
other theories. It would be interesting to evaluate whether these models 
have characteristics useful for late time acceleration, and possibly an 
improvement over standard quintessence potentials. 

In Sec.~\ref{sec:model} we lay out the basic of the models and generalize 
them to a family of models. Section~\ref{sec:results} addresses the 
attractor behavior and presents numerical results for the dark energy 
equation of state evolution. We compare generalized $\al$-models to 
standard quintessence in Sec.~\ref{sec:compare} and discuss their 
advantages, summarizing and mentioning future work in Sec.~\ref{sec:concl}.

\section{$\al$-Models} \label{sec:model} 

The $\al$-model as written for inflation has a non-canonical kinetic 
term and a potential, with Lagrangian density \cite{14123797} 
\bea 
{\mathcal L}&=&\sqrt{-g}\,\left[\frac{1}{2}M_P^2 R-\frac{\al}{(1-\varphi^2/6)^2}\,\frac{1}{2} (\partial\varphi)^2\right.\notag\\ 
&\quad&\left.-\al f^2\left(\frac{\varphi}{\sqrt{6}}\right)\right]\ , 
\eea 
where $M_P$ is the Planck mass, $\al$ is a parameter and $\al f^2$ is the 
potential function. The Starobinsky model corresponds to $\alpha=1$. 

Through a field redefinition 
\be 
\phi=\sqrt{6\al}\,\tanh^{-1}(\varphi/\sqrt{6}) \ , 
\ee 
the kinetic term becomes canonical and the potential function 
\be 
V(\phi)=\al f^2\left(\tanh\frac{\phi}{\sqrt{6\al}}\right) \ . 
\ee 
Note that the function $f$ is not wholly arbitrary since the field 
redefinition breaks down at $\phi=\infty$. 

For example, if we tried to make a standard quintessence model be 
described by an $\al$-model, this 
is not generally possible. Consider dark energy with a constant equation 
of state (pressure to energy density) ratio $w$. This is given by a 
potential \cite{constw} 
\be 
V\sim \sinh^{-2(1+w)/|w|}(\phi-\phi_\star) \qquad [{\rm constant}\ w] \ . 
\ee 
and is accommodated by a function 
\be 
f(x)\sim \left(\frac{x}{\sqrt{1-x^2}}\right)^{-(1+w)/|w|} \ , 
\ee 
where $x=\tanh\,(\phi/\sqrt{6\al})$. 
However this function vanishes at $\phi=\infty$ ($x=1$), and furthermore 
blows up at $\phi=0$. 

For inflation, two functional forms have been used, the T-model 
\be 
f(x)=cx \ , 
\ee 
and the Starobinsky form (with the Starobinsky model having $\al=1$) 
\be 
f(x)=c\,\frac{x}{1+x} \ . 
\ee 
Note that they are equivalent at small $x$, and because $x\propto\phi$ 
at small $\phi$ then near the origin both models give quadratic potentials. 
If we were to take $f(x)\propto x^{p/2}$ for $x\ll1$ then we could match 
onto any monomial potential $\phi^p$ (see, e.g., \cite{14123797}). However 
here we will instead generalize to interpolating and extrapolating these 
two models, to study the deviations from quadratic behavior (but see 
Sec.~\ref{sec:compare}). Our 
generalized $\al$-model takes 
\be 
f(x)=c\,\frac{x}{(1+x)^n} \ . \label{eq:genf}
\ee 
The quantity $c$ scales the amplitude of the potential, and can be 
fixed by requiring a certain dark energy density today; that is, $c$ 
is effectively equivalent to $\Omega_{\phi,0}$, the present dark energy 
density in units of the critical density. The parameter $\al$ scales 
the field value $\phi$.  

\begin{figure}[htbp!]
\includegraphics[width=\columnwidth]{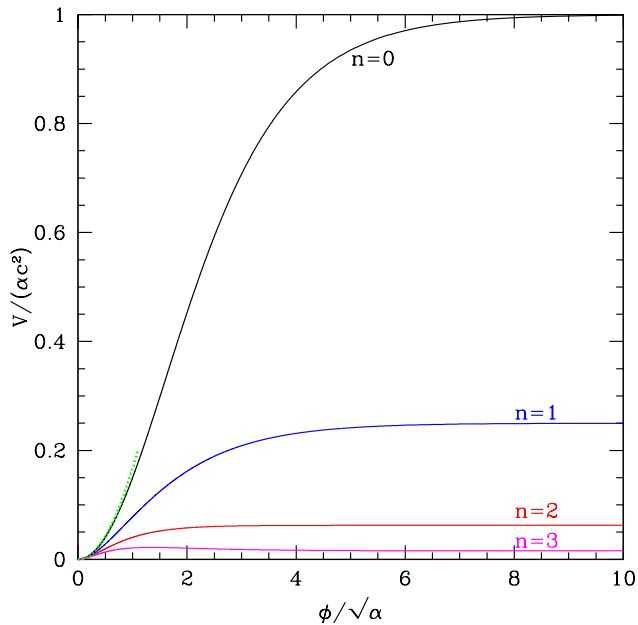} 
\caption{The potential is plotted as a function of the field $\phi$ in 
units of $\sqrt{\al}$, for various values $n$. At small field values 
$\phi\ll\sqrt{\al}$ the $\al$-model resembles the quadratic potential 
$V\propto\phi^2$ (short, dotted green curve), while at large values it 
flattens to an uplifted plateau. 
} 
\label{fig:vphi} 
\end{figure}

As $x\to1$, its maximum value, the potential goes to a constant. Basically 
it plateaus at a level $V_p=\al c^2\,2^{-2n}$. Thus the generalized 
$\al$-model potential looks like a quadratic potential at small $\phi$, 
out to a width $\phi\approx\sqrt{\al}$, and then flattens to a plateau 
at large $\phi$. Figure~\ref{fig:vphi} shows this potential for values 
of $n=0$-3, and the comparison to a standard quadratic potential at 
small $\phi$. (The potential is symmetric about $x=0$ but we show only 
the positive half since the dynamics is symmetric.)

\section{Evolution and Attractors} \label{sec:results} 

In the $\al$-attractor family of inflation theories, the word 
``attractor'' refers to their common values for the scalar perturbation 
power spectrum tilt $n_s$ and tensor to scalar power ratio $r$ in the 
limit $\alpha\to0$, while for large $\al$ the models resemble monomial 
potentials $\phi^p$ \cite{13110472}. Here we are interested in the dynamics 
of the scalar field, and its equation of state evolution, in particular 
whether it gives rise to late time cosmic acceleration. 

From Fig.~\ref{fig:vphi} we can already expect the answer: if the field 
rolls into the quadratic region of the potential, then it should act like 
dark energy from a quadratic potential at late times, while if it remains 
on the plateau then it should act like a cosmological constant. Looking 
at this a bit more carefully, we see that the potential for large $\phi$ 
is 
\be 
V(\phi\gg\sqrt{\al})\approx \al c^2\,2^{-2n}\,
\left[1-2(2-n)\,e^{-2\phi/\sqrt{6\al}}\right] \ . \label{eq:philg} 
\ee 
This is basically an uplifted exponential potential. The exponential 
potential was one of the original dark energy models \cite{wett88}. 
Moreover, the quadratic potential is in the thawing class of dark energy, 
moving away from a cosmological constant state, while the exponential 
potential is in the freezing class, moving toward a cosmological constant. 
(See \cite{caldlin,paths} for discussion of these two main classes of 
dark energy.) 

Thus these $\al$-models appear to conjoin in a way these two classes. 
Furthermore, because the transition depends on the value of $\al$, we 
see that large $\al$ moves the plateau (and hence freezing behavior) to 
larger and larger $\phi$, essentially shrinking the plateau to a point 
maximum and making the potential look like a hilltop potential 
\cite{hilltop}. 

In fact, while these behaviors effectively hold, 
there is a formal de Sitter fixed point at $w=-1$ only for a particular 
range of $n$ (though for other $n$ the field has nearly this behavior, 
for tens of e-folds or more, as we discuss below). 
Therefore we use the terminology $\al$-model rather than 
$\al$-attractor, in the dynamical sense. However, in the inflationary 
sense of a common limit for large $\al$ models, and for small $\al$ models, 
there are attractors as we show in Sec.~\ref{sec:compare}. 

\begin{figure}[htbp!]
\includegraphics[width=\columnwidth]{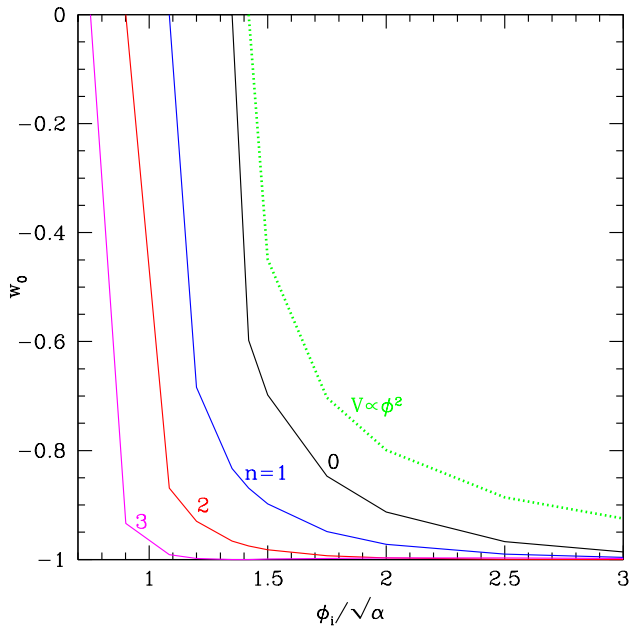} 
\caption{The present value of the equation of state ratio $w_0$ is 
plotted as a function of the initial field value $\phi_i$ in units of 
$\sqrt{\al}$, for various values of $n$ (solid curves). For contrast, 
this relation is also shown for a quadratic potential (dotted green curve, 
where the x-axis is simply $\phi_i$). 
For a given value $\phi_i$, the $\al$-model can achieve $w_0$ closer to 
$-1$ (as preferred by observations). Conversely for a given bound on 
the deviation of $w_0$ from $-1$, the $\al$-model can be less fine tuned 
in the initial field value. 
} 
\label{fig:phii} 
\end{figure}

We numerically solved the scalar field equations of motion and 
Friedmann equations for the background expansion to verify these 
behaviors. Indeed, for initial field values $\phi_i\ll\sqrt{\al}$ 
the field gradually thaws and rolls down the potential, growing from 
equation of state ratio $w=-1$ to less negative values. It 
eventually (perhaps in the far future) oscillates around the quadratic 
minimum, giving a time averaged equation of state ratio $\langle w\rangle=0$. 
However, it can produce a long (if temporary) period of cosmic acceleration. 
Figure~\ref{fig:phii} shows the present equation of state ratio $w_0$ as 
a function of $\phi_i$ for various values of $n$.

For $\phi\gg\sqrt{\al}$, one might suspect that since the plateau is 
not perfectly flat, but tilted upward, that the field could slide down 
into the minimum. However, since the field kinetic energy is rapidly 
damped away in an accelerating universe, this is essentially negligible. 
Indeed, traditional skating fields \cite{paths} where $V=0$ have kinetic 
energy vanishing as $a^{-6}$, where $a$ is the cosmic expansion factor; 
fields on the plateau could be called elevated skaters. 
For example, for $n=0$, where Fig.~\ref{fig:vphi} shows the plateau 
begins at $\phi>5\sqrt{\al}$, the equation of state evolves from 
$w_0=-0.9995$ today to only $w(a=7.5)=-0.9988$ for $\phi_i=5\sqrt{\al}$. 
If $\phi_i=20\sqrt{\al}$, 
then the field rolls only $\Delta\phi/M_P=6\times 10^{-7}$. 

(One might imagine stranger situations, in which the field starts near 
the minimum but with initial velocity away from it high enough to reach 
the plateau, or starting on the plateau and shooting through the minimum 
to reach the $x<0$ plateau. However, these are basically moot because the 
rapid Hubble expansion at early times redshifts away any large kinetic energy, 
requiring extreme fine tuning for these situations to occur.) 

Our generalization shows further interesting properties. 
Figure~\ref{fig:vphi} illustrates that the plateau begins at smaller $\phi$ 
as $n$ increases. Indeed, for $n=2$ and $\phi_i=3\sqrt{\al}$, the equation 
of state evolves from $w_0=-0.9999$ to $w(a=7.5)=-0.9998$. Notably, for 
$n>2$ we have a true attractor since the potential has a maximum on the 
plateau (at $x=1/(n-1)$), and slopes down, rather than up, to its asymptotic 
value as seen 
from Eq.~(\ref{eq:philg}). For example, for 
$n=3$ and $\phi_i>1.35\sqrt{\al}$ the 
asymptotic future state is de Sitter, $w=-1$. 

\begin{figure}[htbp!]
\includegraphics[width=\columnwidth]{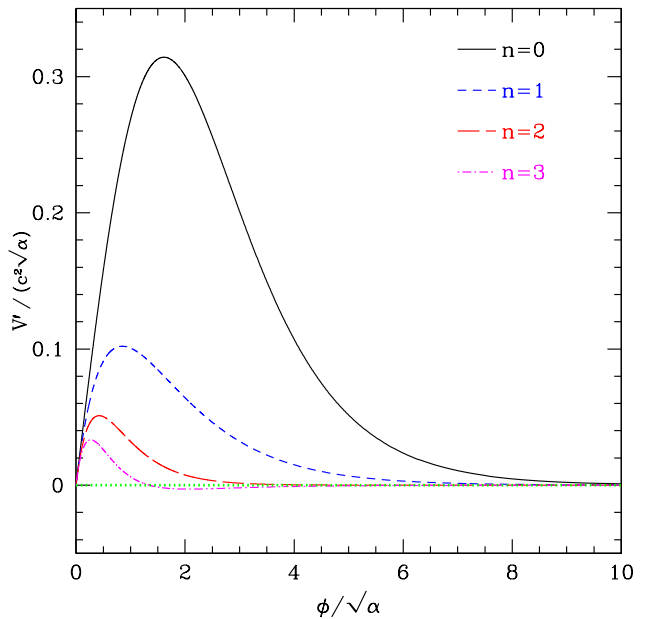} 
\caption{The first derivative of the potential is plotted as a function of 
the field $\phi$ in units of $\sqrt{\al}$, for various values $n$ 
(increasing from top to bottom curves). Note that all derivatives 
approach zero at large field values (since the potential plateau is 
asymptotically flat). 
} 
\label{fig:vp} 
\end{figure}

The potential slope $V'=dV/d\phi$ is given by 
\be 
V'=\sqrt{\frac{2\al}{3}}\,c^2\,\frac{x(1-x)[1+(1-n)x]}{(1+x)^{2n}} \ . 
\ee 
Figure~\ref{fig:vp} shows the slope for various values of 
$n$. We see that the slope rapidly approaches zero for large $\phi$, making 
the plateau quite flat. Recall from the Klein-Gordon equation of motion 
for the scalar field that once the slope is negligible then the field quickly 
freezes in place. For example, for $n=1$ and $\phi_i=5\sqrt{\al}$, the 
field rolls less than $\Delta\phi/M_P=0.06$ in its entire history. 
Thus even when $n\le2$ and there is no formal de Sitter attractor, one 
can still have $w\approx-1$ for a long time.

\section{Relation to Quintessence Models} \label{sec:compare} 

At small $\phi$, the generalized $\al$-model reduces to the quadratic 
potential, 
\be 
V(\phi\ll\sqrt{\al})\approx \frac{c^2}{6\al}\,\phi^2 \ . 
\ee 
That is a thawing dark energy model, and to keep the present 
equation of state ratio $w_0$ near enough to $-1$ to be compatible with 
observational constraints, the initial position of the field requires 
some fine tuning. In particular, starting the field too close to the 
minimum means that it will never achieve a present dark energy density 
in units of the critical density $\Omega_{\phi,0}=0.7$. For example, 
in order to reach this requires $\phi_i>1.42\,M_P$, while to also keep 
$w_0<-0.9$ requires $\phi_i>2.65\,M_P$. 

Now we consider this situation for the $\al$-model. From Fig.~\ref{fig:phii} 
we see the fine tuning can be ameliorated considerably. For $n=0$ ($n=3$) 
the minimum field value to achieve $\Omega_{\phi,0}=0.7$ and $w_0<-0.9$ is 
$\phi_i>1.95\,M_P$ ($0.9\,M_P$) for $\al=1$. 
The conditions to reach $w_0$ closer to 
$-1$ are even more softened, since once the $\al$-model field is on the 
potential plateau then $w$ can easily be extremely close to $-1$. Again, 
the plateau effectively stretches out hilltop potentials, such as from 
pseudo-Nambu goldstone boson or axion fields, or Higgs symmetry breaking, 
vastly increasing the initial conditions delivering $w\approx-1$. 

We can now further generalize the $\al$-model by taking 
\be 
V(x)=\al c^2\,\frac{x^p}{(1+x)^{2n}} \ , 
\ee 
(where before we restricted to $p=2$). This now acts at small $\phi$ 
like any chosen monomial potential $\phi^p$, while retaining the plateau 
at larger field values. True attractors to $w=-1$ can now be achieved 
when the field is at values $x>p/(2n-p)$, requiring $n>p$. For example, 
rather than matching 
the quadratic potential near the minimum we can use the axion monodromy 
potential $\phi^{2/3}$, i.e.\ $p=2/3$. With $n=1$, say, this has a 
late time de Sitter 
attractor for $x>0.5$ or $\phi_i>1.35\,\sqrt{\al}$. 

Let us explore the attractor behavior in the inflationary sense of an 
$\al$-attractor, where there is a common behavior for observables for 
some limit of $\al$. When $\al\gg1$, all models regardless of $n$ will look 
like the $V\propto\phi^p$ case, e.g.\ a quadratic potential for our baseline 
of $p=2$. Conversely, when $\al\to0$, all models will see a plateau potential. 
In particular, when $n>p$ and the field is beyond the potential maximum at 
$x=p/(2n-p)$, then all models are attracted to the de Sitter state. 

\begin{figure}[htbp!]
\includegraphics[width=\columnwidth]{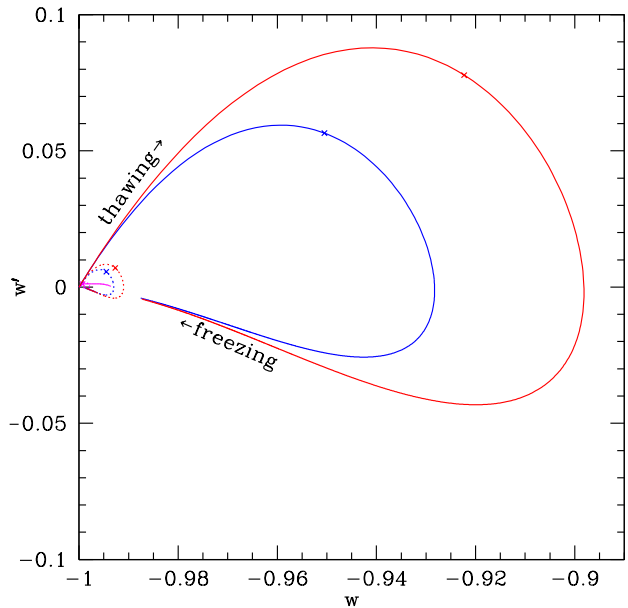} 
\caption{The dynamics in the $w-w'$ phase space transitions between 
thawing and freezing dark energy, with an attractor behavior both 
toward a cosmological constant and toward common dynamics as $\al\to0$. 
Models with $p=2/3$, 1, and 2 (red, blue, magenta, or from outer to inner) 
are shown, for $\al=1$ 
(solid curves) and $\al=0.1$ (dotted curves). The present is denoted by 
x's and the curves are stopped at $a=220$ for clarity. Note the $p=2$, 
$\al=0.1$ case has not quite turned around since it is very close to the 
potential maximum. 
} 
\label{fig:varypa} 
\end{figure}

Figure~\ref{fig:varypa} shows the dynamics in the $w-w'$ phase space, 
where $w'=dw/d\ln a$, for three values of $p$, fixing $n=3$ and taking 
$\phi_i=1.5\,\sqrt{\al}$. For $\al=1$, the dynamics are well 
separated, though with common general characteristics. Note that they all 
start off as thawing dark energy, evolving 
along the canonical thawing behavior of 
$w'=3(1+w)$, and even at present (denoted by x's) they lie in the thawing 
region, at roughly $w'=1+w$. However in the future they turn around and 
convert to freezing behavior, heading along an attractor toward the de Sitter 
point $w=-1$, $w'=0$. For small $\al$, these characteristics persist 
but curves for all $p$ stay much closer to the ultimate cosmological constant 
behavior.

\begin{figure}[htbp!]
\includegraphics[width=\columnwidth]{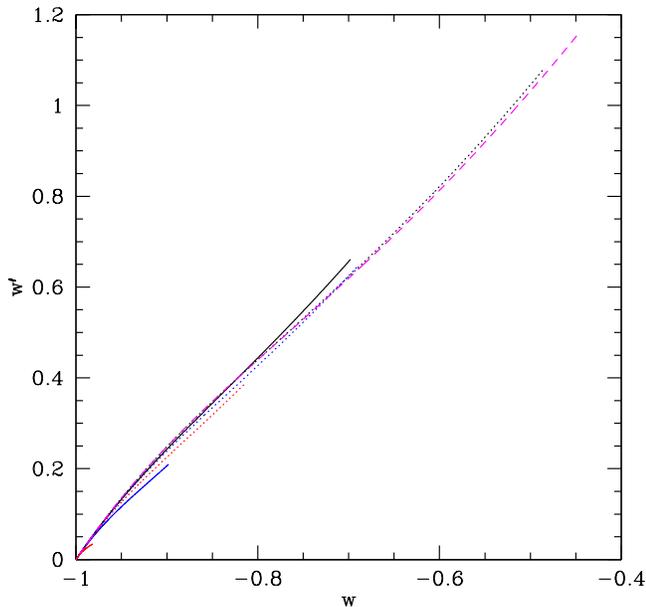} 
\caption{The dynamics in the $w-w'$ phase space increasingly resembles a 
quadratic potential (dashed, long magenta curve) as $\al\to\infty$. 
Models with $n=0$, 1, and 2 (black, blue, red, or from longest to shortest 
curves for each line type) are shown, for $\al=1$ (solid curves) and 
$\al=10$ (dotted curves). For $\al=10$, the curves, and the endpoints at 
$a=1$, move noticeably closer to the quadratic result. 
} 
\label{fig:varyna} 
\end{figure}

Returning to the baseline ($p=2$) case, if we expand the $\al$-model 
potential for small $\phi/\sqrt{\al}$, we see 
\be 
V(\phi\ll\sqrt{\al})\approx \frac{c^2}{6\al}\,\phi^2 
\left(1-\frac{2n}{\sqrt{6\al}}\,\phi+{\mathcal O}(\phi^2/\al)\right) \ . \label{eq:anh}
\ee 
So there is an attractor behavior for large $\al$ where for any $n$ the 
potential acts like a simple quadratic potential. Figure~\ref{fig:varyna} 
illustrates this behavior. (An analogous result will hold for any $p$.)

\begin{figure}[hbp!]
\includegraphics[width=0.96\columnwidth]{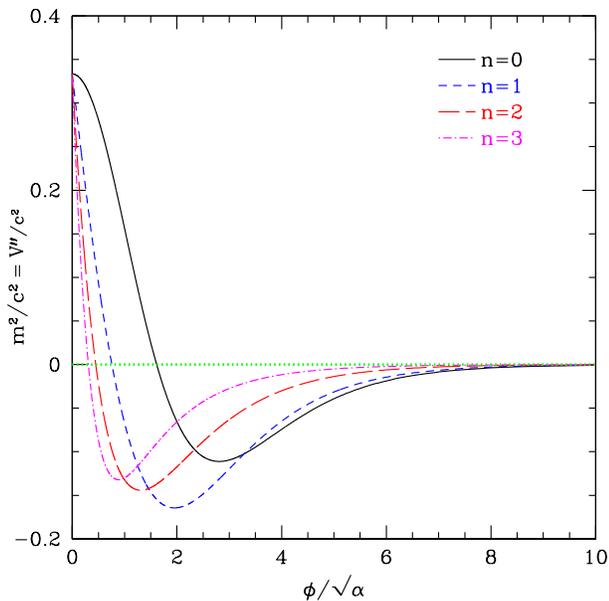} 
\caption{The second derivative of the potential, giving the effective 
mass squared, is plotted as a function of the field $\phi$ in 
units of $\sqrt{\al}$, for various values $n$ (increasing from top to 
bottom curves at small $\phi$). Note that all derivatives 
approach zero at large field values (since the potential plateau is 
asymptotically flat). 
} 
\label{fig:vpp} 
\end{figure}

As an interesting additional point, Eq.~(\ref{eq:anh}) shows that away from 
the minimum the potential becomes anharmonic. 
That is, the potential slope becomes shallower, just as for hilltop 
potentials. We know these have negative mass squared terms near the 
maximum (also see \cite{08051748,11063335,11081793} for other 
anharmonic instabilities), and we find this as well for the $\al$-model. 
Specifically, 
\bea 
m^2=V''&=&\frac{c^2}{3}\,\frac{1-x}{(1+x)^{2n}}\,\left[1+x(1-4n)\right.\\ 
&\quad&\left.-x^2(3-n-2n^2)-x^3(3-5n+2n^2)\right] \ .\notag 
\eea 
Note that the mass vanishes for $\phi\gg1$, as do all the derivatives 
of the potential; the plateau becomes flat. Figure~\ref{fig:vpp} 
exhibits the mass squared as a function of $\phi$ for various $n$.

For hilltop potentials the spinodal, or tachyonic, instability 
when the mass squared is negative is not a major concern
\cite{0001168,0511543,14126879}, 
but for $\al$-models it does exist 
for a wider range of field values. This raises the possibility of interesting 
phenomenology, such as clustering dark energy. Here we consider only the 
background, zero momentum modes, but will investigate this further in 
future work.

\section{Conclusions} \label{sec:concl} 

We investigated the properties of $\al$-attractors, as recently 
highlighted for inflation, as dark energy for late time acceleration 
instead. Such models have several interesting properties, combining 
aspects of the thawing and freezing classes of dark energy together. 
By generalizing the usual $\al$-attractors we derive models that have 
a true de Sitter attractor, as well as ones that have a metastable 
acceleration with equation of state ratio lingering near $w\approx-1$. 

The dynamics near the potential minimum, where the $\al$-model looks 
like a quadratic potential, is found to be less fine tuned that a standard 
quadratic potential, especially for values of $w$ consistent with observations. 
Further from the minimum, the model resembles a stretched hilltop model, 
with elements of an exponential potential as well. 

A family of theories can be defined as one varies the value of $\al$. As 
$\al\to0$, the field sees predominantly a plateau, with a slight slope 
up or down, depending on the value of the generalized parameter $n$. This 
determines whether there is a true de Sitter attractor or not, but either 
way $w\approx-1$ for many e-folds of expansion. For $\al\gg1$, the potential 
increasingly resembles a quadratic potential, or any monomial $\phi^p$ in a 
further generalization. 

We note the $\al$-model has anharmonic properties, and will have negative 
effective mass squared in some regions. Either of these can induce 
clustering in the scalar field, possibly leading to interesting effects. 
While here we concentrate on the background field, future work will explore 
the possibility of observable signatures of these fluctuation properties.

\acknowledgments 

This work has been supported by DOE grant DE-SC-0007867 and the Director, 
Office of Science, Office of High Energy Physics, of the U.S.\ Department 
of Energy under Contract No.\ DE-AC02-05CH11231. 


\end{document}